\documentclass[fleqn,12pt,twoside]{article}
\usepackage{espcrc1}

\usepackage{graphicx}
\usepackage[figuresright]{rotating}

\title{Multiple sources or late injection of short-lived $r$-nuclides in the early solar system?} 
\author{N. Dauphas\address[EFI]{Department of the Geophysical Sciences, Enrico Fermi Institute, and Chicago Center for Cosmochemistry, The University of Chicago, 5734 South Ellis Avenue, Chicago IL 60637, USA} 
}
       
\begin{document}

% typeset front matter
\maketitle
\begin{abstract}
Comparisons between the predicted abundances of short-lived $r$-nuclides (${\rm ^{107}Pd}$, ${\rm ^{129}I}$, ${\rm ^{182}Hf}$, and ${\rm ^{244}Pu}$) in the interstellar medium (ISM) and the observed abundances in the early solar system (ESS) conclusively showed that these nuclides cannot simply be derived from galactic chemical evolution (GCE) if synthesized in a unique stellar environment. It was thus suggested that two different types of stars were responsible for the production of light and heavy $r$-nuclides. Here, new constraints on the ${\rm ^{244}Pu/^{238}U}$ production ratio are used in an open nonlinear GCE model. It is shown that the two $r$-process scenario cannot explain the low abundance of ${\rm ^{244}Pu}$ in the ESS and that this requires either that actinides be produced at an additional site (A-events) or more likely, that ${\rm ^{129}I}$ and ${\rm ^{244}Pu}$ be inherited from GCE and ${\rm ^{107}Pd}$ and ${\rm ^{182}Hf}$ be injected in the ESS by the explosion of a nearby supernova. 
\end{abstract}

\section{Remainder ratios}
Extinct radionuclides were present in the early solar system but have now decayed below detection levels. Their prior presence can be inferred from abundance variations of their daughter nuclides in meteorites. When discussing extinct radionuclides, it is convenient to use the remainder ratio, which is the ratio of the true abundance of the radioactive nuclide to the abundance it would have if it were stable \cite{clayton88,dauphas03}, $\Re=N(\tau)/N(\tau \rightarrow \infty)$, where $\tau$ is the mean-life. Two different approaches can be used to estimate the remainder ratio at solar system formation. The remainder ratio in the ESS can be derived from measurements of ESS materials and nucleosynthesis theory,
\begin{equation}
\Re_{ess}=R/P,
\end{equation}
where $R$ is the ratio of the unstable nuclide to a neighbor stable nuclide synthesized by the same process and $P$ is the associated production ratio. Given a GCE model, it is also possible to calculate the remainder ratio in the ISM at solar system formation ($4567.2\pm 0.6$ Ma for condensation of the first solids \cite{amelin02}). The simplest GCE scenario is the linear closed-box model with instantaneous recycling approximation. This model assumes that ($i$) the Galaxy formed from low metallicity gas that was turned into stars with a rate proportional to the gas density -{\it linear}, ($ii$) that the Galactic disk did not exchange gas with the surrounding medium -{\it closed-box}, and ($iii$) that there was no delay between accretion of gas in stars and return of the nucleosynthetically enriched gas to the ISM -{\it instantaneous recycling}. Under these circumstances, it is straightforward to show that the remainder ratio of a short-lived nuclide in the ISM at 4.5 Gy BP is simply $\tau/{\rm T}$, where T is the presolar age of the galactic disk ($8.7\pm 1.5$ Gy \cite{chaboyer01}). However, this simple model fails to explain important astronomical observations, notably the G-dwarf metallicity distribution. A possible solution to the G-dwarf problem is that the galactic disk accreted low-metallicity gas for an extended period of time. Clayton \cite{clayton88} obtained an analytical expression for the remainder ratio in the framework of a parameterized linear infall model. More recently, Dauphas {\it et al.} \cite{dauphas03} improved over this model by incorporating nonlinearity in the star formation rate (${\rm d}\Sigma _g/{\rm d}t=-\omega \Sigma _g^n$, with $n\simeq 1.4$). In the later model, the rate of infall is parameterized as a truncated gaussian with standard deviation equal to the mean. When infall of low metallicity gas is taken into account, the remainder ratio in the ISM at solar system formation is,
\begin{equation}
\Re_{ism}=\kappa \tau/{\rm T},
\label{eq2}
\end{equation}
 where $\kappa=2.7\pm 0.4$ ($\kappa=1$ would correspond to the closed-box model) \cite{clayton88,dauphas03}. There may have been a delay between isolation of the presolar molecular cloud core from fresh nucleosynthetic inputs and condensation of the first solids in the ESS. This is taken into account by introducing a free decay interval, $\Delta$. The remainder ratio in the ESS is related to the remainder ratio in the ISM through, $\Re_{ess}=\exp (-\Delta /\tau)$. Eliminating $\tau$ in the previous two equations, it follows that there must be a relationship between $\Re _{ism}$ and $\Re _{ess}$,
\begin{equation}
\Re_{ess}=\Re_{ism}\exp \left(-\frac{\kappa \Delta}{{\rm T}\Re _{ism}}\right).
\end{equation}
Irrespective of their mean-lives, if extinct $r$-radionuclides were derived from GCE with the same free decay interval, they should lie on a unique curve parameterized by the previous equation for a specific value of $\Delta$. Wasserburg {\it et al.} \cite{wasserburg96} and Qian {\it et al.} \cite {qian98} did a similar analysis. They concluded that ${\rm ^{129}I}$ and possibly ${\rm ^{107}Pd}$ would require a more extended isolation than ${\rm ^{182}Hf}$ and ${\rm ^{244}Pu}$  ($\kappa=1$ in Fig. 1). This discrepancy led them to suggest that two kinds of $r$-process events were responsible for the nucleosynthesis of neutron-rich nuclides. Low (L) and high (H) frequency events would have synthesized light (${\rm ^{129}I}$) and heavy (${\rm ^{182}Hf}$ and ${\rm ^{244}Pu}$) $r$-process nuclides, respectively. This important work received renewed attention with the observations of $r$-process enriched low metallicity stars. Sneden {\it et al.} \cite{sneden00} analyzed the halo star CS 22892-052 and found that low mass $r$-process elements such as Y and Ag were deficient compared to expectations based on heavier elements. More recently, Hill {\it et al.} \cite{hill02} determined the abundances of U, Th, and Eu in another halo star, CS 31082-001. The age of this star based on the Th/Eu ratio would be lower than the age of the solar system and is inconsistent with an independent age determination based on the U/Th ratio of $14.0\pm 2.4$ Ga. These observations point to multiple $r$-processes (possibly as many as 3).

\section{Plutonium-244}
Plutonium-244 has a mean life of 115.416 My. Its abundance in the ESS normalized to ${\rm ^{238}U}$ is ${\rm ^{244}Pu/^{238}U=6.8\pm 1.0\times 10^{-3}}$ \cite{hudson88}. For most extinct $r$-nuclides, it is possible to estimate production ratios based on a decomposition into $r$ and $s$-processes of the daughter nuclides \cite{meyer00}.  In contrast, the ${\rm ^{244}Pu/^{238}U}$ production ratio must be estimated based entirely on theoretical grounds. The main difficulty with this approach is that there are no immediate stable neighbor nuclides to anchor the models. Goriely and Arnould \cite{goriely01} estimated uncertainties in the production of actinides. Among the various models that they list, only those that yield ${\rm ^{235}U/^{238}U}$ and ${\rm ^{232}Th/^{238}U}$ production ratios consistent with solar system abundances are retained \cite{hill02,goriely01}. The ${\rm ^{244}Pu/^{238}U}$ production ratio is thus estimated to be $0.53\pm 0.36$ (cases 1, 3, 4, 7-9, 13, 20, 22, and 30 of \cite{goriely01}). The ratio $\Re ^{244}_{ess}/\Re ^{238}_{ism}$ is therefore $1.28\pm 0.89 \times 10^{-2}$. The remainder ratio of ${\rm ^{238}U}$ at solar system formation can be estimated in the framework of infall GCE models to be ${\Re^{238}_{ess}\simeq 0.71}$ (0.53 in the closed-box model) \cite{clayton88,dauphas03}. The remainder ratio of ${\rm ^{244}Pu}$ is therefore $\Re^{244}_{ess}=9.1\pm 6.3\times 10^{-3}$. The remainder ratio in the ISM based on Eq. \ref{eq2} is $\Re^{244}_{ism}=3.58\pm 0.81 \times 10^{-2}$ (nonlinear infall model with $\kappa=2.7$ \cite{dauphas03}). The free decay interval for ${\rm ^{244}Pu}$ ($\Delta=\tau \ln \left(\Re^{244}_{\rm ism}/Re^{244}_{\rm ess}\right)$) is thus estimated to be $158\pm 85$ My (Fig. 1). The production ratios, ESS abundances, ISM, and ESS remainder ratios of other short-lived $r$-nuclides are compiled in Table 1 \cite{meyer00,dauphas03}. 

\begin{table}[htbp]
{\scriptsize
\caption{Extinct $r$-radionuclides in the ESS (${\rm ^{107}Pd}$ might have an $s$-process origin).}
\begin{tabular}{lcccccc}
\hline
\hline
Nuclide & ${\tau}$ (My) & Norm. & $R$ & $P$ & ${\Re ^i_{ess}}$ & ${\rm \Re ^i_{ism}}$\\ %& $\Delta$ (My) \\
\hline

Palladium-107 & 9.378 & ${\rm ^{110}Pd^r}$ & $5.8\pm 1.2 \times 10^{-5}$ & $\sim 1.36$ & $4.3\pm 0.9 \times 10^{-5}$ & $2.91\pm 0.66\times 10^{-3}$ \\%& $40\pm 3$\\
Iodine-129 & 22.650 & ${\rm ^{127}I^r}$ & $1.25\pm 0.21 \times 10^{-4}$ & $\sim 1.45$ & $8.6\pm 1.5 \times 10^{-5}$ & $7.03\pm 1.6\times 10^{-3}$ \\%& $100\pm 7$\\
Hafnium-182 & 12.984 & ${\rm ^{177}Hf^r}$ & $3.70\pm 0.58 \times 10^{-4}$ & $\sim 0.81$ & $4.57\pm 0.72\times 10^{-4}$ & $4.03\pm 0.92 \times 10^{-3}$ \\%& $28\pm 3.6$\\
Plutonium-244 & 115.416 & ${\rm ^{238}U}$ & $6.8\pm 1.0 \times 10^{-3}$ & $5.3\pm 3.6 \times 10^{-1}$ & $9.1\pm 6.3\times 10^{-3}$ & $3.58\pm 0.81\times 10^{-2}$ \\%& $158\pm 84.1$\\
\hline
\end{tabular}\\[2pt]
}
 The superscript $r$ refers to the $r$-process component of the cosmic abundances \cite{arlandini99}. $R$ is the ratio observed in the ESS, $P$ is the production ratio, $\Re^i_{ess}$ is the remainder ratio in the ESS calculated as $R/P$ (except for ${\rm ^{244}Pu}$, for which the normalizing ratio is unstable and the remainder ratio must be corrected for $\Re^{238}=0.71$), $\Re^i_{ism}$ is the remainder ratio in the ISM as obtained from open nonlinear GCE modeling \cite{dauphas03}.
\label{table1nd}
\end{table}

\begin{figure}[htbp]
\begin{center}
\includegraphics[width=11.5cm]{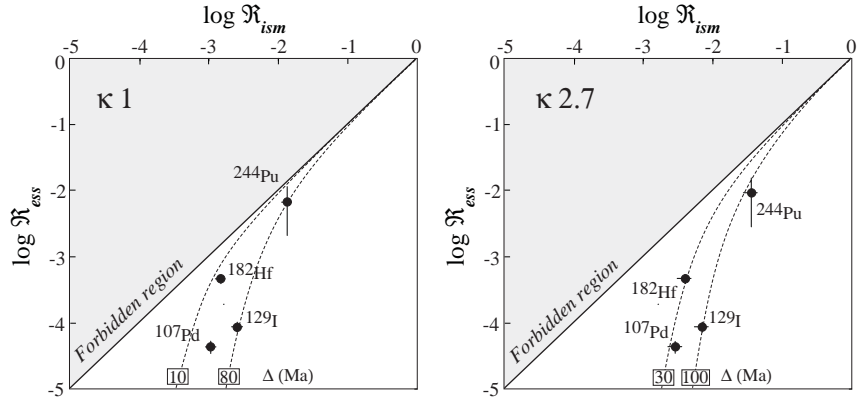}
\end{center}
\caption{Remainder ratios in the ESS and the ISM for the closed box model (left panel, $\kappa=1$, $\Re^{238}_{ess}=0.53$) and the open nonlinear GCE model (right panel, $\kappa=2.7$, $\Re^{238}_{ess}=0.71$, table 1 \cite{dauphas03}), for a variety of free decay interval $\Delta$. As illustrated, if infall of low metallicity gas is taken into account, ${\rm ^{244}Pu}$ cannot be ascribed to the same origin as ${\rm ^{182}Hf}$.}
\end{figure}

\section{Implications on the multiple $r$-processes scenario}
The multiple $r$-process scenario as advocated by Wasserburg {\it et al.} \cite{wasserburg96} and Qian {\it et al.} \cite{qian98} assumes that light and heavy $r$-process nuclides were produced in different stellar environments. Plutonium-244 should therefore follow ${\rm ^{182}Hf}$, which requires a free-decay interval of $\sim 30$ My. Although affected by large uncertainty (stemming mainly from uncertainty in the production ratio), the free-decay interval of ${\rm ^{244}Pu}$ is closer to that of ${\rm ^{129}I}$ than it is to that of ${\rm ^{182}Hf}$ (Fig. 1, right panel). The dichotomy between light (${\rm ^{129}I}$) and heavy (${\rm ^{182}Hf}$ and ${\rm ^{244}Pu}$) $r$-nuclides may not be valid. This may indicate that actinides were synthesized in a different stellar environment (A-events). There is indeed evidence for decoupling between actinides and other $r$-process nuclides in low-metallicity stars \cite{hill02}. Another possible scenario is that the nuclides that are overabundant in the ESS compared to GCE expectations (${\rm ^{107}Pd}$ and ${\rm ^{182}Hf}$ but also ${\rm ^{26}Al}$, ${\rm ^{36}Cl}$, ${\rm ^{41}Ca}$, and ${\rm ^{60}Fe}$) were injected in the presolar molecular cloud core by the explosion of a nearby supernova (SN) that might have triggered the protosolar cloud into collapse. In the most recent version of the SN pollution model \cite{meyer04}, it is assumed that only a fraction of the stellar ejecta is efficiently injected in the nascent solar system. For a stellar mass of 25 ${\rm M_{\odot}}$, an injection mass cut of 5 ${\rm M_{\odot}}$, and a time interval of 1 Ma between the SN and incorporation in the ESS, the abundances of ${\rm ^{26}Al}$, ${\rm ^{41}Ca}$, ${\rm ^{60}Fe}$, and ${\rm ^{182}Hf}$ are successfully reproduced while the abundances of ${\rm ^{36}Cl}$ and ${\rm ^{107}Pd}$ are slightly overproduced but this may reflect uncertainties in ESS abundances and input physics \cite{meyer04}. Because the multiple $r$-processes scenario requires many stellar sources for explaining all extinct radionuclides that cannot be produced by GCE or irradiation in the ESS while the injection scenario requires only one, the principle of Occam's razor would favor the SN pollution model.

{\bf Acknowledgements.} I thank A.M. Davis, T. Rauscher, B. Marty, and L. Reisberg for fruitful discussions. This work was supported by NASA grant NAG5-12997 to A.M. Davis.


\begin{thebibliography}{9}
{\scriptsize
\bibitem{clayton88} D.D. Clayton, Mon. Not. R. Astron. Soc. 234 (1988) 1.
\bibitem{dauphas03} N. Dauphas, T. Rauscher, B. Marty, L. Reisberg, Nucl. Phys. A 719 (2003) 287c.
\bibitem{amelin02} Y. Amelin, A.N. Krot, I.D. Hutcheon, A.A. Ulyanov, Science 297 (2002) 1678.
\bibitem{chaboyer01} B. Chaboyer, in: T.V. Hippel, C. Simpson, N. Manset (Eds.), Astrophysical Ages and Time Scales, ASP Conference Series 245, Astronomical Society of the Pacific, San Francisco, 2001, p. 162.
\bibitem{wasserburg96} G.J. Wasserburg, M. Busso, R. Gallino, Astrophys. J. 466 (1996) L109.
\bibitem{qian98} Y.-Z. Qian, P. Vogel, G.J. Wasserburg, Astrophys. J. 494 (1998) 285. 
\bibitem{sneden00} C. Sneden, J.J. Cowan, I.I. Ivans, G.M. Fuller, S. Burles, T.C. Beers, J.E. Lawler, Astrophys. J. 533 (2000) L139.
\bibitem{hill02} V. Hill, B. Plez, R. Cayrel, T.C. Beers, B. Nordstr\"om, J. Andersen, M. Spite, F. Spite, B. Barbuy, P. Bonifacio, E. Depagne, P. Fran\c{c}ois, F. Primas, Astron. Astrophys. 387 (2002) 560.
\bibitem{hudson88} G.B. Hudson, B.M. Kennedy, F.A. Podosek, C.M. Hohenberg, Lunar Planet. Sci. XIX (1988) 547.
\bibitem{meyer00} B.S. Meyer, D.D. Clayton, Space Sci. Rev. 92 (2000) 133.
\bibitem{goriely01} S. Goriely, M. Arnould, Astron. Astrophys. 379 (2001) 1113.
\bibitem{arlandini99} C. Arlandini, F. K\"appeler, K. Wisshak, R. Gallino, M. Lugaro, M. Busso, O. Straniero, Astrophys. J. 525 (1999) 886.
\bibitem {meyer04} B.S. Meyer, L.-S. The, D.D. Clayton, M.F. El Eid, Lunar Planet. Sci XXXV (2004) \#1908.


}
\end{thebibliography}
\end{document}